\documentclass[aip,reprint]{revtex4-1}

\usepackage[author={DLG}]{pdfcomment}

\usepackage{xcolor}
\usepackage{graphicx}
\usepackage[english]{babel}
\usepackage{mathtools}
\usepackage{bm}
\usepackage{bbm}

\begin{document}

\title{Optimal control gradient precision trade-offs: application to fast
generation of DeepControl libraries for MRI}

\author{Mads Sloth Vinding \textsuperscript{$\dagger$,}}
\email[]{msv@cfin.au.dk}
\thanks{\textsuperscript{$\dagger$} These two authors contributed equally}
\affiliation{Center of Functionally Integrative Neuroscience (CFIN), Department of Clinical Medicine, Faculty of Health, Aarhus University, Denmark}

\author{David L. Goodwin\textsuperscript{$\dagger$,}}
\email[]{david.goodwin@partner.kit.edu}
\affiliation{Institute for Biological Interfaces 4 - Magnetic Resonance, Karlsruhe Institute for Technology (KIT), Karlsruhe, Germany}
\affiliation{Chemistry Research Laboratory, University of Oxford, Mansfield Road, Oxford, UK}

\author{Ilya Kuprov}
\affiliation{School of Chemistry, University of Southampton, Southampton, SO17 1BJ, UK}

\author{Torben Ellegaard Lund}
\affiliation{Center of Functionally Integrative Neuroscience (CFIN), Department of Clinical Medicine, Faculty of Health, Aarhus University, Denmark}

\date{\today}

\begin{abstract}
We have recently demonstrated supervised deep learning methods for rapid generation of radiofrequency pulses in magnetic resonance imaging (\url{https://doi.org/10.1002/mrm.27740}, \url{https://doi.org/10.1002/mrm.28667}). Unlike the previous iterative optimization approaches, deep learning methods generate a pulse using a fixed number of floating-point operations - this is important in MRI, where patient-specific pulses preferably must be produced in real time. However, deep learning requires vast training libraries, which must be generated using the traditional methods, e.g. iterative quantum optimal control methods. Those methods are usually variations of gradient descent, and the calculation of the fidelity gradient of the performance metric with respect to the pulse waveform can be the most numerically intensive step. In this communication, we explore various ways in which the calculation of fidelity gradients in quantum optimal control theory may be accelerated. Four optimization avenues are explored: truncated commutator series expansions at zeroth and first order, a novel midpoint truncation scheme at first order, and the exact complex-step method. For the spin systems relevant to MRI, the first-order truncation is found to be sufficiently accurate, but also up to five times faster than the machine precision gradient. This makes the generation of training databases for the machine learning methods considerably more realistic.
\end{abstract}


\maketitle

\section{Introduction}

Tailored radiofrequency (RF) pulses are used in advanced magnetic resonance imaging (MRI) applications: reduced field-of-view imaging \cite{ma_reduced_2013,maximov_real-time_2015,saritas_dwi_2008}, spectral-spatial selectivity \cite{ma_reduced_2013,vinding_dynamic_2013}, 
as well as imaging with inhomogeneous RF fields ($B_1^+$) \cite{skinner_optimal_2012}, off-resonance effects \cite{stockmann_vivo_2018,stockmann_32-channel_2016,juchem_dynamic_2011,juchem_magnetic_2009} or gradient imperfections \cite{aigner_time_2019}. Optimal control is well suited to generating the necessary RF pulse shapes, including multi-channel pulses, several milliseconds long with microsecond time stepping and arbitrary flip angles \cite{vinding_dynamic_2013,lee_time-optimal_2009,ulloa_chebyshev_2004,vinding_application_2017,vinding_local_2017,xu_designing_2008,grissom_spatial_2006}.

One method, called gradient ascent pulse engineering (GRAPE), uses a piecewise-constant pulse approximation that is well suited for MRI \cite{khaneja_optimal_2005}. Here, the term gradient is in relation to optimization and not to be confused with the magnetic field gradients. GRAPE was initially designed using the gradient ascent method (linear convergence), and then extended to quasi-Newton (super-linear convergence) methods \cite{de_fouquieres_second_2011}, and Newton-Raphson (quadratic convergence) methods \cite{goodwin_modified_2016}. The latter two require very accurate gradients \cite{de_fouquieres_second_2011,goodwin_modified_2016,goodwin_auxiliary_2015,kuprov_derivatives_2009}. Importantly, tailored RF pulses in MRI often need to be computed in real time - the patient is inside the scanner. However, GRAPE uses iterative optimization that has no wall clock time guarantees: for realistic systems, it can become impractically slow.

We have recently proposed a neural network based method for generating RF pulses to alleviate the run time problem \cite{vinding_ultrafast_2019,deepcontrol21,vinding_ISMRM21}. This \emph{DeepControl} framework requires training libraries with hundreds of thousands of RF pulses generated, e.g., with optimal control, but a trained neural network predicts a pulse quickly, and with a hard guarantee on the wall clock time.

These vast libraries require an improvement in our ability to run optimal control simulations and calls for optimization of the numerical efficiency of the GRAPE algorithm, in particular, in the part that deals with the gradient calculation.

In this paper, we report some efficiency improvements in the GRAPE gradient calculation process. We analyze the trade-offs between CPU time and convergence rate for four different levels of approximation to the GRAPE gradient: standard zero and first order approximation, a novel midpoint first order approximation proposed here, and the exact, complex-step gradient. 

\section{Optimal control theory}

The equation of motion of a system of uncoupled spin-1/2 particles, ignoring relaxation, can be formulated as:
\begin{equation}
\dot{\bm{M}}_{}(\bm r, t)=\bm{\Omega}_{}(\bm r, t)\bm{M}_{}(\bm r, t)
\label{BlochEquation}
\end{equation}
where $\bm{M}_{}(\bm r, t)$ is an instantaneous magnetization vector at a spatial location $\bm r$, and $\bm{\Omega}_{}(\bm r, t)$ is the dynamics matrix containing external magnetic fields, defined below. By the nature of this study, we introduce discrete notation from this point. The spacial dimension is separated into $P$ different locations, with $1<p<P$, and for the $p^\text{th}$ location $\bm r_{}^{(p)} = [x_{}^{(p)},y_{}^{(p)},z_{}^{(p)}]^\top$ ($^\top$ is the vector transpose operation).

The control of these dynamics is in the form of a piecewise-constant \emph{control waveform} with a fixed total duration $T$, divided into $N$ time steps of duration $\Delta t = T/N$, denoted a time slice. The magnetization state we are attempting to control, $\bm M_{n}^{(p)} = [M_{x,n}^{(p)},M_{y,n}^{(p)},M_{z,n}^{(p)}]^\top$, will have $N+1$ temporal instances for each spatial position, with an initial state $\bm{M}_{0}^{(p)}$ and a final state $\bm{M}_{N}^{(p)}$. Hence, for the state vectors $0<n<N$, or in other words the $n^\text{th}$ control propagates the system from the $(n-1)^\text{th}$ state to the $n^\text{th}$ state. The task for an optimal control problem is to bring this final magnetization state as close as possible to a desired target state by maximizing the overlap of a target magnetization state, $\bm M_\text{target}^{(p)}$, and the final magnetization state, $\bm{M}_{N}^{(p)}$.

Dynamics of Eq.~(\ref{BlochEquation}) are governed by $\bm{\Omega}$, and the piecewise-constant formulation of this is
\begin{equation}
\bm{\Omega}_{n}^{(p)}\!=\! 
\gamma\!\!\left[
\begin{matrix}
0 &+\bm{G}_n^\top \bm{r}_{}^{(p)} + B_{z}^{(p)} & -B_{y,n}^{(p)}\\
-\bm{G}_n^\top \bm{r}_{}^{(p)} - B_{z}^{(p)} & 0 & +B_{x,n}^{(p)}\\
 +B_{y,n}^{(p)} & - B_{x,n}^{(p)} & 0
\end{matrix}
\right],
\label{eq:Omega}
\end{equation}
which includes the RF fields in the Zeeman rotating frame, $B_{x,n}^{(p)}$ and $B_{y,n}^{(p)}$, the magnetic field gradient, $\bm{G}_n = [G_{x,n},G_{y,n},G_{z,n}]^\top$, the magnetic field inhomogeneity, $B_{z}^{(p)}$, and the gyromagnetic ratio, $\gamma$.

Xu et al.~\cite{xu_designing_2008} have shown how to extend these equations to multiple RF channels, $1<l<N_\text{Tx}$, i.e., parallel transmit (pTx) with complex coil sensitivity patterns, $s_{l}^{(p)}$:
\begin{align}
B_{x,n}^{(p)} =& \sum\limits_{l=1}^{N_\text{Tx}} \left[ \mathrm{Re}\left(s_{l}^{(p)}\right) c_{u,l,n} -\mathrm{Im}\left(s_{l}^{(p)}\right) c_{v,l,n} \right]\\
B_{y,n}^{(p)} =& \sum\limits_{l=1}^{N_\text{Tx}} \left[ \mathrm{Im}\left(s_{l}^{(p)}\right) c_{u,l,n} +\mathrm{Re}\left(s_{l}^{(p)}\right) c_{v,l,n} \right]
\end{align}
Here, $c_{j,l,n}\in\{c_{u,l,n},c_{v,l,n}\}$ are the RF control sequences to be optimized to achieve the aim, and all other parameters are considered fixed. For generality, we introduce the control $c_{j,l,n}$, with the subscript $j$ as reference to either $u$ or $v$, when it is possible to imply to either.

Discrete solutions to Eq.~(\ref{BlochEquation}), in the piecewise-constant approximation, can be written as
 \begin{equation}
\bm M_{n}^{(p)} = \mathbf U_{n}^{(p)} \bm M_{n-1}^{(p)}
\label{eq:BlochSol}
\end{equation}
where $\bm M_{0}^{(p)}$ is a given initial state of the system and $\mathbf U_{n}^{(p)}$ is a time-propagator describing a rotation on the Bloch sphere, which can be calculated with the matrix exponential:
 \begin{equation}
\mathbf U_{n}^{(p)} = \mathrm{e}^{\mathbf \Omega_{n}^{(p)}\Delta t}
\label{eq:U}
\end{equation}

Note, the magnetic field gradient waveform is specified for the entire pulse duration and kept fixed, although it could also be included as a control to be optimized. 

Our performance functional, describing our aim, is the projection of $\bm M_{N}^{(p)}$, onto the target state $\bm M_\text{target}^{(p)}$, summed and normalized over all spatial points:
 \begin{equation}
J = \frac{1}{P} \sum_{p=1}^P {\bm{M}_{N}^{(p)}}^\top \cdot \bm M_\text{target}^{(p)}
\label{eq:J1}
\end{equation}
Thus, $J$ is the quantity we want to maximize and is termed the \emph{fidelity}. Further to this fidelity measure, which is dependent on only the final state of the system and termed the \emph{terminal cost}, an additional term can be included which is termed a \emph{running cost} and depends on the state of the system over all of the pulse duration. The running cost can be regularization terms, to penalize untargeted control behavior, e.g., excess power or jagged waveforms \cite{goodwin_modified_2016}, or boundary constraints. In this work we use simple boundary constraints, which are detailed in \cite{vinding_local_2017}.

\section{Fidelity gradient calculations}

The formal theory of optimal control introduces an \emph{adjoint state} of the system, $\bm L_{n}^{(p)}$, with the detailed derivation published elsewhere \cite{vinding_fast_2012}. This adjoint state of the system can be interpreted as the propagation of the target, backwards in time from $t=t_{N}$ to $t=t_0$, and for a particular time slice this is
\begin{equation}
\bm L_{n-1}^{(p)} =  {\mathbf U_{n}^{(p)}}^\top \bm L_{n}^{(p)}
\label{eq:BlochSolBwd}
\end{equation}
where $\bm L_{N}^{(p)}=\bm M_\text{target}^{(p)}$. The adjoint state is used in calculation of the fidelity gradients, $\nabla J_{u,l,n}^{}$ and $\nabla J_{v,l,n}^{}$, with respect to $c_{u,l,n}$ and $c_{v,l,n}$, enabling the use of gradient-following numerical optimization methods to maximize $J$.

The scope of this study is to cast light on a number of different strategies for calculating $\nabla J_{u,l,n}^{}$ and $\nabla J_{v,l,n}^{}$. Common prerequisites for the different gradient strategies are a forward time-propagation of the magnetization from the initial to the final state for all positions by Eq.~(\ref{eq:BlochSol}), and a backward time-propagation of the adjoint state for all positions by Eq.~(\ref{eq:BlochSolBwd}). 

\subsection{Standard approximate gradients}

Derivatives of functions of matrices, $f(\mathbf \Omega)$, are called \emph{directional derivatives}, or \emph{G\^{a}teaux derivatives}, which are defined in Ref.~\cite{higham_functions_2008} as
\begin{align}
D_{\bm{\Theta}}^{}(f(\bm{\Omega}))\triangleq& \lim_{h\to0}\frac{f(\bm{\Omega}+h\bm{\Theta}) - f(\bm{\Omega})}{h}\nonumber\\
=&\frac{\mathrm{d}}{\mathrm{d} h}\bigg|_{h=0}\!\!\!f(\bm{\Omega}+h\bm{\Theta}) \label{eqn:directionalderivative}
\end{align}
where $\bm{\Theta}$ is the operator matrix for the control in question and elaborated below. 

Eq.~(\ref{eqn:directionalderivative}) is a similar form to the finite difference equation and the Taylor series truncated to first order and should be considered the formal definition of the derivative of the function of a matrix.

For the dynamic system of Eq.~(\ref{eq:U}) we are considering here, the \emph{directional derivative} here can be written as \cite{blanes2009magnus}
\begin{align}
 D_{\bm{\Theta}}^{}(\mathrm{e}^{\bm{\Omega}\Delta t}) & =\Bigg[\int\limits_0^{\Delta t}\!\mathrm{e}^{\bm{\Omega}s}\bm{\Theta}\mathrm{e}^{-\bm{\Omega}s}\mathrm{d} s\Bigg]\mathrm{e}^{\bm{\Omega}\Delta t}\\
 & = \bigg\{\bm{\Theta},\!\!\int\limits_0^{\Delta t}\!\mathrm{e}^{\bm{\Omega}s}\mathrm{d} s\bigg\}\mathrm{e}^{\bm{\Omega}\Delta t}= \bigg\{\bm{\Theta},\frac{\mathrm{e}^{\bm{\Omega}\Delta t}-\bm{\mathbbm{1}}}{\bm{\Omega}} \bigg\}\mathrm{e}^{\bm{\Omega}\Delta t}\nonumber\\
 & = \Bigg[\sum\limits_{r=0}^{\infty}\frac{(\Delta t)_{}^{r+1}}{(r+1)!}\big\{\bm{\Theta},\bm{\Omega}_{}^{r} \big\}\Bigg]\mathrm{e}^{\bm{\Omega}\Delta t}
 \label{eq:dirderiv}
\end{align}
where $\{\bm{A},\bm{B}\}$ denoted the anti-commutator of the matrices $\bm{A}$ and $\bm{B}$. The first order approximation, $r=1$, corresponds to equation (12) in the original GRAPE paper \cite{khaneja_optimal_2005}. For the controls $c_{u,l,n}$ and $c_{v,l,n}$~\cite{xu_designing_2008}, the corresponding control operators are
\begin{align}
 \bm{\Theta}_{u,l}^{(p)} =& \left[
\begin{matrix}
0 & 0 & -\mathrm{Im}\left(s_{l}^{(p)}\right) \\
0 & 0 & +\mathrm{Re}\left(s_{l}^{(p)}\right)\\
+\mathrm{Im}\left(s_{l}^{(p)}\right) & -\mathrm{Re}\left(s_{l}^{(p)}\right) & 0
\end{matrix}
\right]\\
 \bm{\Theta}_{v,l}^{(p)} =& \left[
\begin{matrix}
0 & 0 & -\mathrm{Re}\left(s_{l}^{(p)}\right) \\
0 & 0 & -\mathrm{Im}\left(s_{l}^{(p)}\right)\\
+\mathrm{Re}\left(s_{l}^{(p)}\right) & +\mathrm{Im}\left(s_{l}^{(p)}\right) & 0
\end{matrix}
\right]
\end{align}

In addition, if we optimized the magnetic field gradient waveform, e.g. $G_{g,n}$ with $g = x,y,z$, its control operator is
\begin{equation}
 \bm{\Theta}_{G_g}^{(p)}= \left[
\begin{matrix}
0 & +\gamma g^{(p)} & 0 \\
-\gamma g^{(p)} & 0 & 0 \\
0 & 0 & 0
\end{matrix}
\right], g = x,y,z
\end{equation}

In our vectorized model, Eqs.~(\ref{eq:Omega}) to (\ref{eq:BlochSolBwd}), we consider the zeroth, $r=0$, and first order, $r=1$, gradient approximations of Eq.~(\ref{eq:dirderiv}):
\begin{align}
\nabla J_{j,l,n}^{(\text{STD-0})} =&\frac{\Delta t}{P}\sum_{p=1}^P {\bm{L}_{n}^{(p)}}^\top \Big[ \mathbf \Theta_{j,l}^{(p)} \Big]\bm M_{n-1}^{(p)}+\mathcal{O}(\Delta t)\label{eq:nabJ0}\\
\nabla J_{j,l,n}^{(\text{STD-1})} =&\frac{\Delta t}{P}\sum_{p=1}^P  {\bm{L}_{n}^{(p)}}^\top  \Big[\mathbf \Theta_{j,l}^{(p)} \mathbf U_{n}^{(p)}\Big]\bm M_{n-1}^{(p)}+\mathcal{O}(\Delta t^2)\label{eq:nabJ1}
\end{align}
where the big-$\mathcal{O}$-notation qualifies the error, which depends mainly on the size of the time step \cite{de_fouquieres_second_2011}.

Had we included regularization in Eq.~(\ref{eq:J1}) this would also be present in the gradient terms above \cite{goodwin_modified_2016}. 

The zeroth and first-order gradient approximations are widely used and fast to compute \cite{machnes_comparing_2011}, but for a quasi-Newton method they would quickly corrupt the Hessian estimation and slows the super-linear convergence to linear, especially for long time increments \cite{de_fouquieres_second_2011}. De Fouquieres et al.~\cite{de_fouquieres_second_2011} show how gradients computed to increasing precision improve the performance of quasi-Newton methods. 

\subsection{Midpoint approximate gradients}

In this paper, we propose a \emph{midpoint} variant of the standard approximate gradient to partly overcome the convergence problem of the standard approximate gradients. The midpoint variant was inspired by the central finite-difference gradient, although it has a deep theoretical basis in the \emph{exponential midpoint method} \cite{hochbruck1997krylov,lubich2002integrators,hochbruck2003magnus,blanes2009magnus,hochbruck2010exponential}

The midpoint method for this manuscript is
\begin{align}
\nabla J_{j,l,n}^{(\text{MID})} 
=\frac{\Delta t}{2P}\sum_{p=1}^P
& {\bm{L}_{n}^{(p)}}^\top \Big[ {\mathbf{U}_{n-1}^{(p)}}  \mathbf \Theta_{j,l}^{(p)}\nonumber\\
&+ \mathbf \Theta_{j,l}^{(p)} \mathbf U_{n}^{(p)} \Big]\bm M_{n-1}^{(p)} +\mathcal{O}(\Delta t^2)\label{eq:nabJmid}
\end{align}
which is an average of the $(n-1)^\text{th}$ and $n^\text{th}$ gradient elements; centralizing the derivative to the midpoint of the $(n-1)^\text{th}$ and $n^\text{th}$ time point. The propagator at $t_0$ is the identity matrix so, when expanded, ${\bm{L}_{1}^{(p)}}^\top \mathbf{U}_{0}^{(p)}  \mathbf \Theta_{j,l}^{(p)}\bm M_{0}^{(p)}={\bm{L}_{1}^{(p)}}^\top \mathbf \Theta_{j,l}^{(p)}\bm M_{0}^{(p)}$.

Although this may seem a simple extension to Eq.~(\ref{eq:nabJ1}) without obvious benefit, since both have an error $\mathcal{O}(\Delta t^2)$, a deeper read to the literature shows the $\mathcal{O}(\Delta t^2)$ error term in Eq.~(\ref{eq:nabJ1}) has a constant hidden in the $\mathcal{O}$-notation. The \emph{hidden constant} depends on the time derivatives of $\bm{M}$ and the bounds of the $\bm{\Omega}$ \cite{lubich2002integrators,hochbruck2003magnus,hochbruck2010exponential}. The midpoint method in Eq.~(\ref{eq:nabJmid}) does not suffer from this error dependency \cite{hochbruck1997krylov} and can be important when space discretization is used \cite{blanes2009magnus}. Furthermore, Eq.~(\ref{eq:nabJ0}) and Eq.~(\ref{eq:nabJ1}) are good approximations only when solutions are mildly oscillating \cite{borzi_qc_2017}, which is not expected to be the case for MRI. Although not included in the analysis that follows, the inclusion of gradient control will accentuate this last point, as the gradient controls introduce another highly oscillating part to the Bloch equations -- particularly during the start of the optimization. It should be emphasized that the exponential midpoint method \cite{hochbruck1997krylov,lubich2002integrators,hochbruck2003magnus,blanes2009magnus,hochbruck2010exponential} is designed for approximation of time propagators, which is not the case in this study; time propagators in this study are exact, and the derivatives of time propagators, $\mathbf{D}_{j,l,n}^{(p)}$ defined by Eq.~(\ref{eq:dirderiv}), are approximated using a midpoint method.

\subsection{Exact Gradient}

Floether et al.~\cite{floether_robust_2012} presented a method to compute exact control gradients using an auxiliary matrix approach \cite{loan_computing_1978}, and Goodwin et al.~\cite{goodwin_auxiliary_2015} extended this to the second order derivatives needed for a Hessian. We recently adopted the exact gradient calculation in our optimal control framework, and described it in Ref.~\cite{vinding_local_2017}. Briefly,
\begin{equation}
\nabla J_{j,l,n}^{(\text{EXACT})} = \frac{1}{P}\sum_{p=1}^P {\bm{L}_{n}^{(p)}}^\top \Big[\mathbf{D}_{j,l,n}^{(p)}\Big] \bm{M}_{n-1}^{(p)}
 \label{eq:exact}
\end{equation}

The gradient in Eq.~(\ref{eq:exact}) is exact to machine precision \cite{goodwin_auxiliary_2015}. The essence of this exact approach is the derivation of $\mathbf{D}_{j,l,n}^{(p)}$ in Eq.~(\ref{eq:exact}). This can be obtained from the upper right corner of an auxiliary matrix \cite{loan_computing_1978,floether_robust_2012,goodwin_auxiliary_2015} but for real matrices $\bm{\Omega}_{n}^{(p)}$ and $\bm{\Theta}_{j,l}^{(p)}$, the compact method of \emph{complex-step approximation} \cite{al-mohy_complex_2009} can be used to calculate the directional-derivative, $\mathbf{D}_{j,l,n}^{(p)}$. This method gives the propagator and derivative as the real and imaginary parts of a complex matrix, respectively:
\begin{equation}
 \mathbf{U}_{n}^{(p)} + i \mathbf{D}_{j,l,n}^{(p)} = \exp\left\{ \left[\bm{\Omega}_{n}^{(p)} + i \bm{\Theta}_{j,l}^{(p)} \right]\Delta t\right\},
 \label{eq:complex}
\end{equation}
where $\bm{\Omega}_{n}^{(p)}$ and $\bm{\Theta}_{j,l}^{(p)}$ must be real matrices, which can be the case in the single-spin model. 

Considering any one given spatial point, $p$, the complex-step differentiation involves $3\times3$ complex matrices instead of $6\times6$ real matrices. With that, there is potential to spare memory allocation.  Complex-step differentiation is also known to have better roundoff error tolerance in finite precision arithmetic \cite{al-mohy_complex_2009}.

\subsection{Accuracy tests}

\begin{figure}
\includegraphics[width=8cm, angle=0]{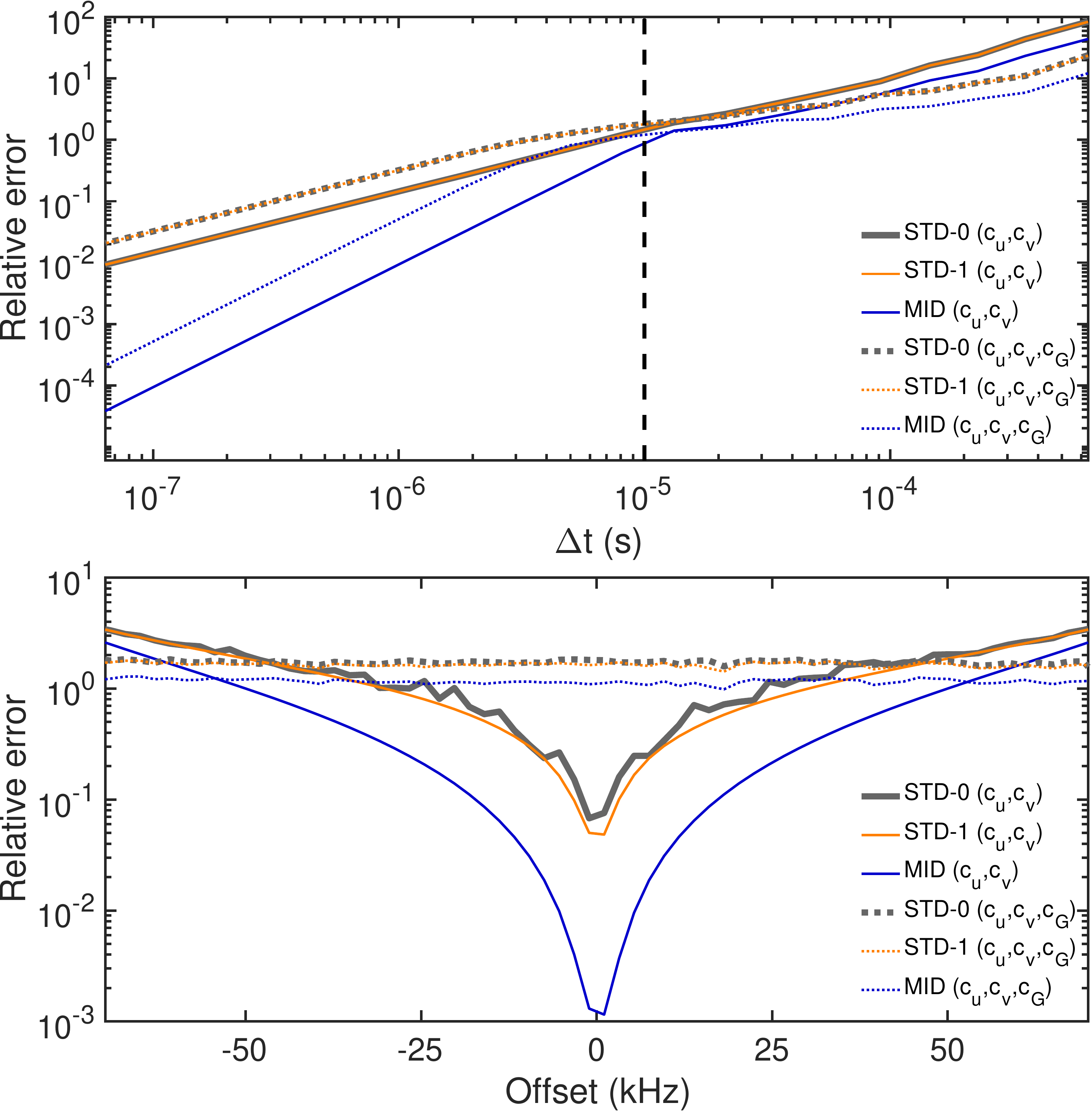} 
\caption{ Relative accuracy of approximate gradients compared to the exact gradient in Eq.(\ref{eq:exact}). Both plots show the relative accuracy of the standard zeroth order (STD-0), Eq.(\ref{eq:nabJ0}), first order (STD-1), Eq.(\ref{eq:nabJ1}), and first order midpoint (MID), Eq.(\ref{eq:nabJmid}), gradients. (Top) Errors as a function of time slice, $\Delta t$, with the dashed vertical line being the $\Delta t$ used in the study. (Bottom) Errors as a function of offset for $\Delta t=10$~$\mu$s, corresponding to the dashed vertical line in the top plot. Both plots show the error when field gradient controls ($c_G$) are used together with RF controls ($c_u, c_v$) with dotted lines, and when it is just RF controls (as in this study) with full lines.}
\label{FigAccuracy}
\end{figure}

The relative accuracy of gradients calculated with the methods in Eq.(\ref{eq:nabJ0}), Eq.(\ref{eq:nabJ1}), and Eq.(\ref{eq:nabJmid}), is shown in Figure \ref{FigAccuracy}. The error in the accuracy is calculated with
\begin{equation}
\xi = \frac{|| \nabla J -  \nabla J_{}^{(\text{EXACT})}||_2}{||\nabla J_{}^{(\text{EXACT})}||_2}
\end{equation}
and averaged over 20 instances of random pulses. Each pulse has a duration of $T=6.39$~ms, with the random distribution between $\pm 1$~kHz for RF controls, $c_u$ and $c_v$, and $\pm 160$~kHz for magnetic field gradient controls, $c_G$. The tests were done firstly (top, Fig.~\ref{FigAccuracy}) with time slice, $\Delta t$, ranging logarithmically in 20 steps from 63.9~ns to 0.639 ms for an offset of $B_{z}^{(p)} = 0$~Hz. Secondly (bottom, Fig.~\ref{FigAccuracy}), the offset was picked randomly in a bandwidth of 160~kHz for a fixed $\Delta t = 10$~$\mu$s. The number of offset points is approximately one point per $2.1$~kHz of bandwidth, and they are spread randomly within the given bandwidth. The initial state of the system is all spins having positive $z$-magnetization, and the target state is all spins having positive $x$-magnetization i.e. a $90^\circ$ rotation around the y-axis of the Bloch sphere.

The accuracy comparison in Figure \ref{FigAccuracy} shows that the MID gradient gives an increase in accuracy, even though the method of calculation in Eq.(\ref{eq:nabJmid}) requires only the insignificant computational cost of a matrix addition per gradient element. The dashed vertical line in the top panel of Figure \ref{FigAccuracy} indicates the time slice, $\Delta t=10$~$\mu$s, used in the results that follow. This time slice is chosen as it is physically relevant to the MRI hardware, but also from approximately this time slice the MID gradient becomes increasingly more effective than the STD-0 and STD-1 gradients for decreasing time slices. The pulse duration, bandwidths, and control amplitudes were based on the settings described in the next section.

\section{Methods}

An optimal control theory framework was implemented in MATLAB (Mathworks, Natick, MA, USA), with hard constraints on the RF pulse amplitudes using the \texttt{fmincon} optimization function supplemented with a performance gradient function and the necessary housekeeping procedures. We focus here on the gradients required by the quasi-Newton option of \texttt{fmincon}, although we are not limited to GRAPE. 

The four gradients types have been implemented\cite{data} in the blOCh framework used in Refs.~\cite{maximov_real-time_2015,vinding_fast_2012,vinding_dynamic_2013,vinding_application_2017,vinding_local_2017,vinding_ultrafast_2019,deepcontrol21}.

A common feature for these calculations is that each direct or adjoint state must be known at a given time slice, $n$, and/or at the adjacent time slice, $n-1$. Conducting full magnetization forward propagation, and adjoint state backward propagation solves this issue. In principle, however, the magnetization state can be calculated as gradient element calculations proceed, i.e., $\mathbf{U}_{n}^{(p)} $ can also be extracted from Eq.~(\ref{eq:complex}), in the same way as propagator recycling implemented for Hessian calculations \cite{goodwin_advanced_2017}, possibly with efficient matrix caching \cite{goodwin_modified_2016}.

The most efficient code for parallelizing the propagation and gradient elements is specific to computing architecture. The two most obvious parameters considered are the amount of CPU cores available, and the storage size needed for all matrices from forward and backward propagation. Our computations were performed on a 28-CPU core, Intel Xeon Gold 5120, 2.2 GHz workstation with 384 GB of RAM.

The EXACT gradient of Eq.~(\ref{eq:complex}) was assessed regarding parallelization efficiency. An L-curve analysis is shown in Figure S1 in the Supporting Information.  The EXACT gradient was on the present workstation most efficient with parallelization on around 23 out of the 28 available workers. Accordingly, the reported computation times for the EXACT gradient are with 23-fold parallelization. The STD-0, STD-1 and MID gradients are efficient without parallelization, and we report computation times for the fastest (non-parallelized) implementations we could develop.

We evaluated optimization performance with the four different gradients by computing single-channel 2DRF pulses with a library of total 150 different flip-angle (FA) maps of three different target categories: 
\begin{enumerate}
\item 50 binary images, where ten randomly placed points were dilated and merged to a single binary shape with no holes. These are denoted BW (black-white), see Figure S2.
\item 50 normalized gray-scale images made binary with a 0.5-threshold, denoted GrBW, see Figure S3.
\item50 normalized gray-scale images, denoted Gr (Gray), see Figure S4.
\end{enumerate}
 
Images were taken from the \emph{ImageNet} database \cite{krizhevsky_imagenet_2017}, and the nominal target FA was 30$^{\circ}$. The target magnetization $x$- and $z$-component levels were dictated by the image intensity levels multiplied with the nominal FA. The target $y$-component was 0. The initial magnetization was longitudinal. The Gr category represents our ability to induce spatially variant FAs, while the binary categories are two random ad hoc ways to obtain shaped excitation patterns, mimicking anatomic regions of interest.

The gradient waveform accommodating the spatial selectivity of the 2DRF pulses formed an inward 16-turn spiral in the excitation k-space (full Nyquist sampling) with a duration of 6.39 ms, a field of excitation of 25 cm, and limited to 40 mT/m amplitude and 180 T/m/s slew rate \cite{lee_time-optimal_2009}. The common time slice for RF and field gradients was 10 $\mu$s. RF pulses were constrained below 1 kHz (nutation rate) amplitude using the boundary option implemented in \texttt{fmincon}. The spatial grid was 64$\times$64 with a mask formed by a centrally placed super ellipse, which reduced the number of spatial points to 2919. The quasi-Newton method with STD-0, STD-1, MID and EXACT gradients was limited with the algorithm termination conditions of reaching 100 iterations or a minimum functional/step-size convergence tolerance of $1\cdot10^{-6}$. 

\section{Results}

For all three categories (BW, GrBW and Gr), all optimizations with approximate gradients converged. The optimizations with EXACT gradients were stopped in 32\% of all cases by the iteration limit. 

\begin{figure*}
\includegraphics[width=17cm, angle=0]{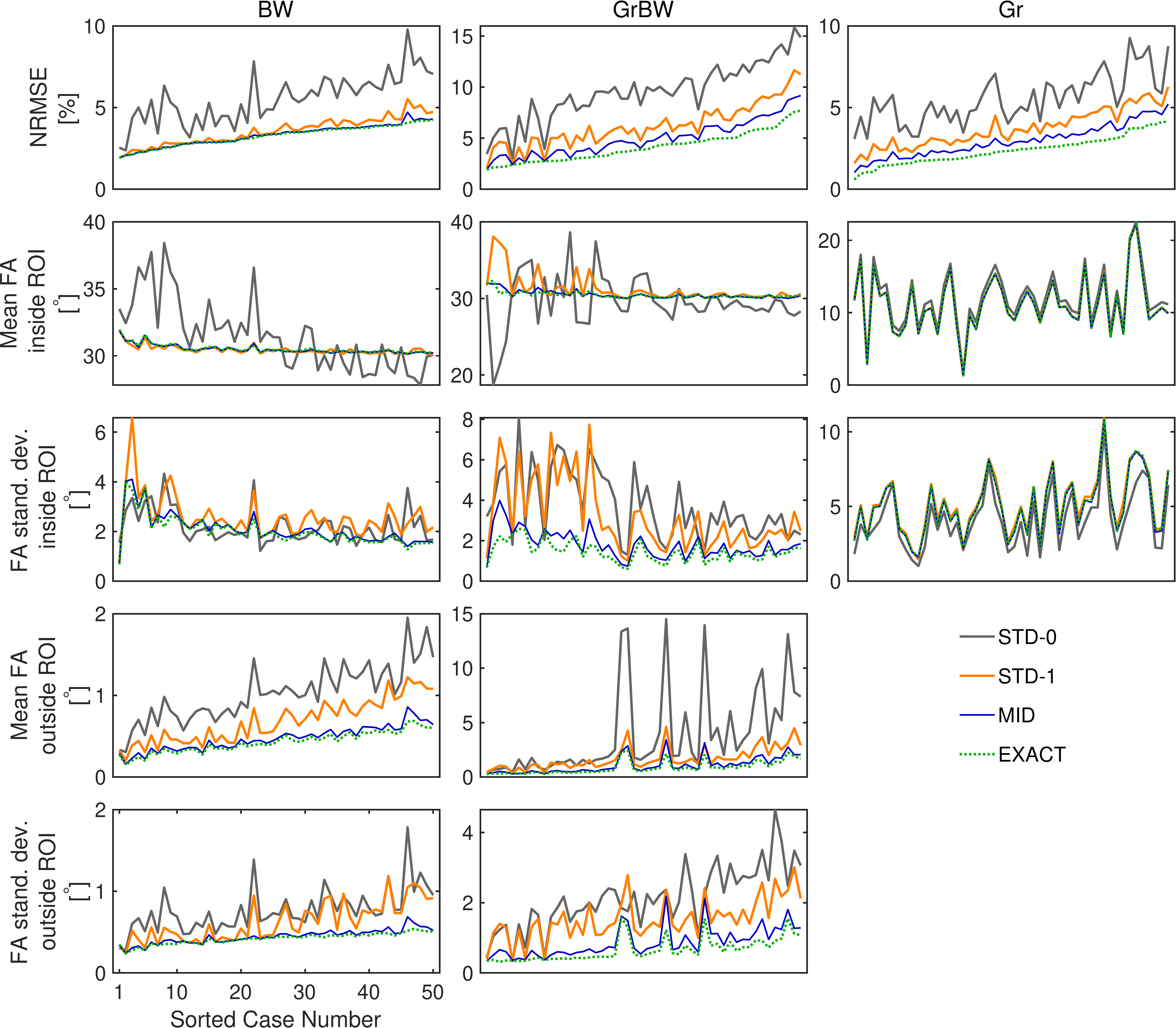}
\caption{Plots of NRMSE and FA statistics for the different target categories: BW (left column), GrBW (center column), Gr (right column). The case numbers are sorted based on the NRMSE values from the optimization, i.e., the error of the Bloch simulated magnetization with respect to the target magnetization, based on the EXACT gradient results. For the Gr target category, the entire pattern is considered as inside the ROI, thus there are no plots for outside areas. In the same category, the expected mean and standard deviation FA inside the ROI is not expected or supposed to be $30^\circ$ and $0^\circ$, respectively, but may vary according to the gray levels.}
\label{FigStats}
\end{figure*}

Figure \ref{FigStats} shows the NRMSE and FA statistics from all Bloch simulations. The cases are sorted by the NRMSEs of the EXACT gradient) within each target category. All FA maps are shown in Figure S2-4 in the Supporting Information. 

\begin{table*}
 \caption{
 Iterations and computation times (mean $\pm$ standard deviation). Diagonal entries with two mean $\pm$ standard deviation pairs correspond to total use of a given method, i.e., computation times include gradient and approximate Hessian estimation, step size search, hard-constraints, and all other internal routines of the interior-point algorithm used within MATLAB's \texttt{fmincon} function. 
 The upper triangular entries include three mean $\pm$ standard deviation pairs (from top to bottom): 1) the number of iterations it takes for the method in the column to outperform the method in the row considering the NRMSE value; 2) the approximate time it takes the column method to overtake the row method; 3) the approximate time the row method was overtaken by the column method. The approximate times are estimated by  the ratio of iterations needed for crossing and the total number of iterations multiplied by the total computations times for all iterations. 
 Target categories (BW, GrBW and Gr) are pooled for the measures in this Table.
 }
 \label{Tab:stats} 
 \begin{tabular}{| c || c | c | c | c || c |}
 \hline \hline
Gradient Method &  STD-0  & STD-1 & MID & EXACT &  \\\hline
STD-0 		& $6.5 \pm 0.8 $ 		& $5.0 \pm 0.4 $  		& $4.9 \pm 0.4 $  		& $4.8 \pm 0.4 $ 		& Iterations \\
   &      $11.9 \pm 0.6 $     &     $5.6 \pm 1.0 $      &     $5.7 \pm 0.7 $      &       $19.8 \pm 1.7 $    &     Seconds \\
   &     &     $9.3 \pm 1.0 $      &     $9.1 \pm 0.7 $      &       $9.0 \pm 1.7 $    &     Seconds \\\hline
STD-1  		&  & $14.0 \pm 6.3 $  		& $7.0 \pm 1.1 $  		& $6.6 \pm 1.1 $ 		& Iterations\\
  &  & $14.7 \pm 2.9 $  		& $7.9 \pm 0.9 $  		& $27.0 \pm 4.5 $ 		 & Seconds \\
  &  &  & $7.7 \pm 1.2 $  		& $7.2 \pm 1.1 $ 		 & Seconds \\\hline
MID   		&  &   		& $13.1 \pm 2.5 $  		& $8.1 \pm 1.5 $ 		& Iterations\\
   &        &     & $14.9 \pm 1.6 $   &        $33.1 \pm 5.9 $ 		 & Seconds \\
   &        &     &    &        $9.2 \pm 1.1 $ 		 & Seconds \\\hline
EXACT  		&  &   		&   		& $73.5 \pm 25.4 $ 		& Iterations\\
 		&  		&  		&   		& $303.8 \pm 109.7 $ 		 & Seconds \\\hline
 \end{tabular}
 \end{table*}

Table \ref{Tab:stats} lists iteration and computation time statistics (mean and standard deviation) for the three target categories (BW, GrBW and Gr) pooled together. Of interest (see the diagonal entries), the MID and STD-1 gradients spend on average roughly the same amount of iterations and time to finish, around 14 iterations in 15 seconds, respectively. The EXACT method spends on average 74 iterations and nearly 300 seconds to finish. 

\begin{figure*}
\includegraphics[width=17cm, angle=0]{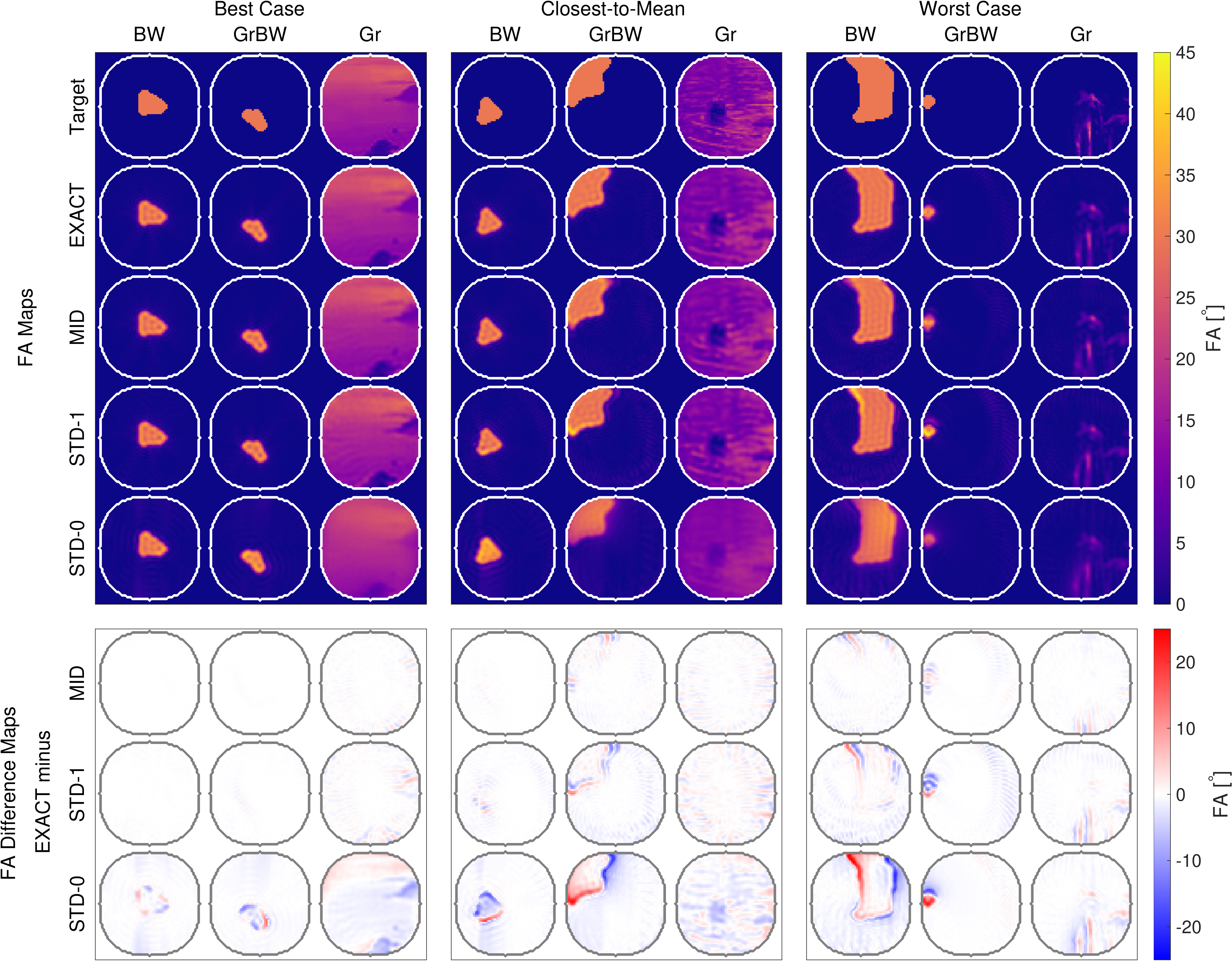}
\caption{FA maps of the best (left), closest-to-mean (middle) and worst (right) cases, when contrasting the NRMSE of the FA maps of the MID gradient with respect to the EXACT gradient. The corresponding STD-0 and STD-1 results are shown as well. Differences are shown in the lower part. Only pulses from converged solutions were considered for this Figure.}%
\label{FigBC_MnC_WC_FA_maps}
\end{figure*}
 
In comparison (see the off diagonal entries), the MID gradient is not outperformed by the EXACT gradient until it reaches 8 iterations or 33 seconds, which is equivalent to around 9 seconds for the MID gradient. However, as shown in Figure \ref{FigStats} the MID and EXACT gradients have near identical performance. This is supported FA maps displayed in Figure \ref{FigBC_MnC_WC_FA_maps}. 

Judged from the NRMSE of FA maps (MID against EXACT), we compared the best, closest-to-mean, and worst cases for converged pulses. As reference, we also show the results from the corresponding STD-0 and STD-1 gradients, as well as the difference maps, highlighting the increasing errors using less precise gradients.

\begin{figure}
\includegraphics[width=7cm]{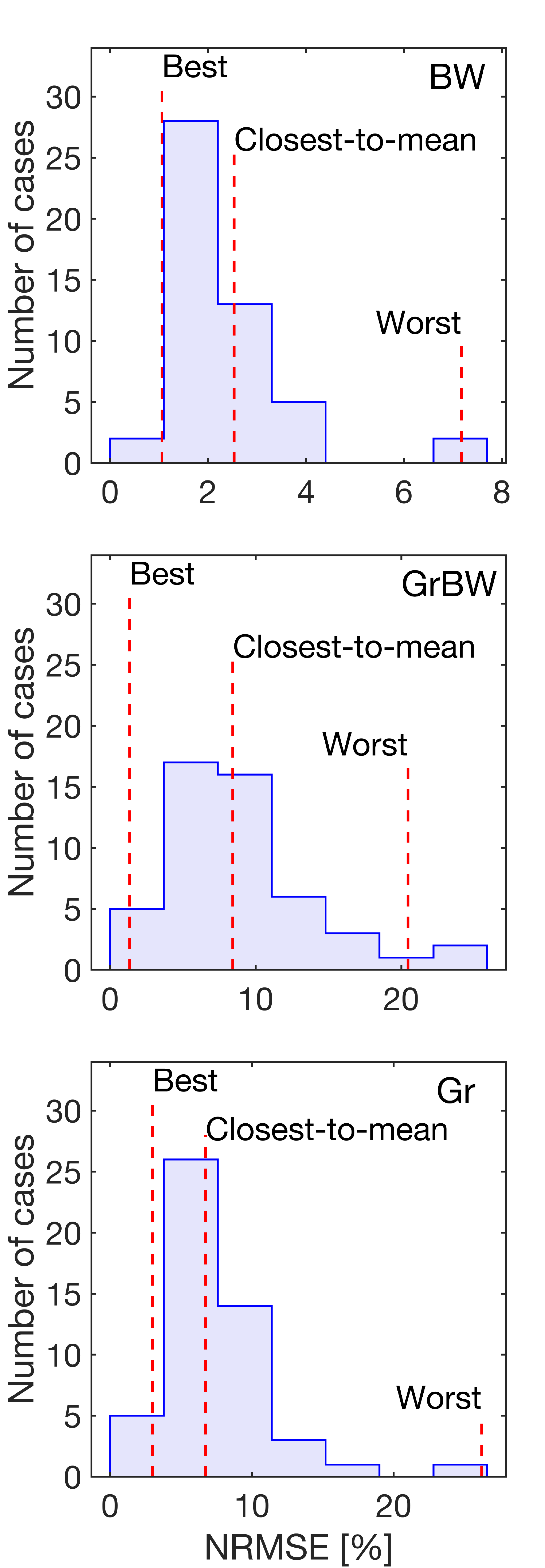}
\caption{NRMSE histograms contrasting the MID gradient against the EXACT gradient, for the BW (top), GrBW (middle) and Gr (bottom) target categories. The best, closest-to-mean and worst case examples shown in Figure \ref{FigBC_MnC_WC_FA_maps} are herein signified by the dashed lines. The histograms contain all cases, i.e., also pulses terminated by reaching the maximum number of iterations, why the histogram edges extend below and beyond the best and worst case levels, respectively.
}%
\label{FigHistograms}
\end{figure}

Figure \ref{FigHistograms} shows the NRMSE histograms of all pulses, where the right skew reveals that the mean NRMSE is close to the best cases, and the worst cases are rare.

\section{Discussion}

We have presented a midpoint first order gradient approximation, denoted MID, with a similar computation time to a standard first order gradient approximation computation, and the accuracy similar to the exact, complex-step gradient for typical MRI settings.

The average time per iteration for the present settings increase by a factor of around 1.01 going from the standard first order to the midpoint gradient method, however by a factor of five going from the standard first order to the exact, complex-step gradient bearing in mind the exact, complex-step gradient exploit 23-fold parallelization. With four CPU cores, typically available on a laptop, we estimate the complex-step gradient will further require a factor of two of computation time.

The standard and the midpoint gradients are efficient without parallelization, and easily vectorized. Several designs for the approximate gradients have been tested for the RF pulse design. However, we could not find any implementation, where either gradient becomes faster from parallelization for the present application: the bottleneck of transferring data to / from the workers dominates the workload. For single-channel RF optimization and the spatial grid size ($P$) and number of time-steps ($N$) as reported above, the standard gradients are slightly faster being run in a for-loop over time steps, rather than as a more vectorized version. There was no significant difference between the for-loop and vectorized versions of the midpoint gradient.
As a side-note, when running multi-channel RF optimization (not the case herein), both the standard and midpoint gradients benefit from the vectorized version and the time difference between them diminishes. However, there remains a for-loop over the number of RF channels currently.
Accordingly, approximate gradient computation times were reported herein for the fastest, non-parallelized implementations we have found.

Optimization speed is inherently important for in vivo experiments, but also for generating AI training libraries. 

Considering the vast training libraries required for \emph{DeepControl} \cite{vinding_ultrafast_2019,deepcontrol21}, the non-parallelized, vectorized computation of the standard and midpoint approximate gradients is not a problem, when we exploit all available CPU cores with an \emph{outer parallelization} over many independent pulses.  

It is beyond the scope of this work to investigate how the qualities of the different gradients influence the \emph{DeepControl} framework, but we do expect infidelities of a given library to propagate through to the final trained neural network, and that predicted pulses never perform better than the representative training library. This study leaves the question of what is more important, fast computation time or pulse perfection, as an additional option for the user. We also propose the midpoint gradient method for rapidly producing a good initial guess that may be handed over to a slower, exact gradient method for finalizing.

While it is possible to further improve the accuracy of the standard or midpoint approximate gradient by lowering the time slice, this may not be feasible on realistic hardware. The accuracy of the midpoint gradient corresponds roughly to that of the standard 0$^\text{th}$ order gradient, STD-0, with a halved time slice. 

This was found by running the same experiments (data not shown) with the STD-0 for a time slice of 5 $\mu$s instead of 10 $\mu$s.

The choice for target categories reflect our previous \cite{vinding_ultrafast_2019,deepcontrol21,vinding_ISMRM21} and future \emph{DeepControl} experiments, which will be described in a subsequent publication. We acknowledge that routines operating in the so-called small-flip-angle regime pose a robust alternative to optimal control in terms of speed and accuracy due to the linear Fourier relation existing between the pulse waveform and the excitation pattern.  As shown in Ref.~\cite{vinding_ultrafast_2019}, we trained networks for both small and large flip angles, with Fourier and optimal control based algorithms generating the libraries. However, we note that the optimal control framework we use enables any arbitrary FA (and flip phase) for each individual spatial and spectral position. It is due to our \emph{DeepControl} experiments that we chose in this study to target a nominal FA of 30$^\circ$ (fair to say in the small-flip-angle regime) together with optimal control, which was not strictly necessary for this FA, but chosen for several other reasons, one being because our \emph{DeepControl} experiments benefit from hard pulse constraints, which we have so far only implemented in our optimal control framework. Yet, hard constraints do exist in various other pulse designs \cite{aigner_time_2019,hoyos-idrobo_variant_2014}. It is however important to mention that the individual gradient computation times do not change for other, say, large flip angles. 

\section{Conclusion}

Pulse waveform optimization gradient calculation using a midpoint first order approximation was evaluated and found to provide a significant efficiency improvement. This illustrates clearly that the current trend of using exact gradients in all situations should be revised - tailored approximations are much cheaper and can be just as accurate. Approximate gradients have been observed to make quasi-Newton pseudo-Hessian less efficient at accelerating convergence, but we have not encountered this phenomenon here, likely because the terminal fidelity requirements were less stringent than they are in quantum technology applications.

The midpoint gradient method has been tested herein for 2DRF pulses, with the intent to create a large neural network training library, but it is also applicable to other pulse types, e.g. multidimensional parallel transmit pulses of any target flip angle, and other control types, e.g. field gradients and novel multi-channel shim waveforms \cite{vinding_instant_2020}. These applications have a large number of channels, and would benefit significantly from the fast gradient computation and the relatively high accuracy of the midpoint approximate gradient method outlined in this paper.

\begin{acknowledgments}
We thank Villum Fonden, Eva og Henry Fraenkels Mindefond, Harboefonden, and Kong Christian Den Tiendes Fond.
\end{acknowledgments}

\section*{Data Availability}

The data that supports the findings of this study are available within the article and its supplementary material. Our code repository at https://github.com/madssakre will contain the proposed gradient implementations including capabilities for parallel transmit.

\bibliographystyle{apsrev4-1}
\bibliography{GradPrecBib}

\clearpage

\onecolumngrid
\appendix

\section{Computation Times}
 \begin{figure*}[h!]
 \includegraphics[width=\textwidth, angle=0]{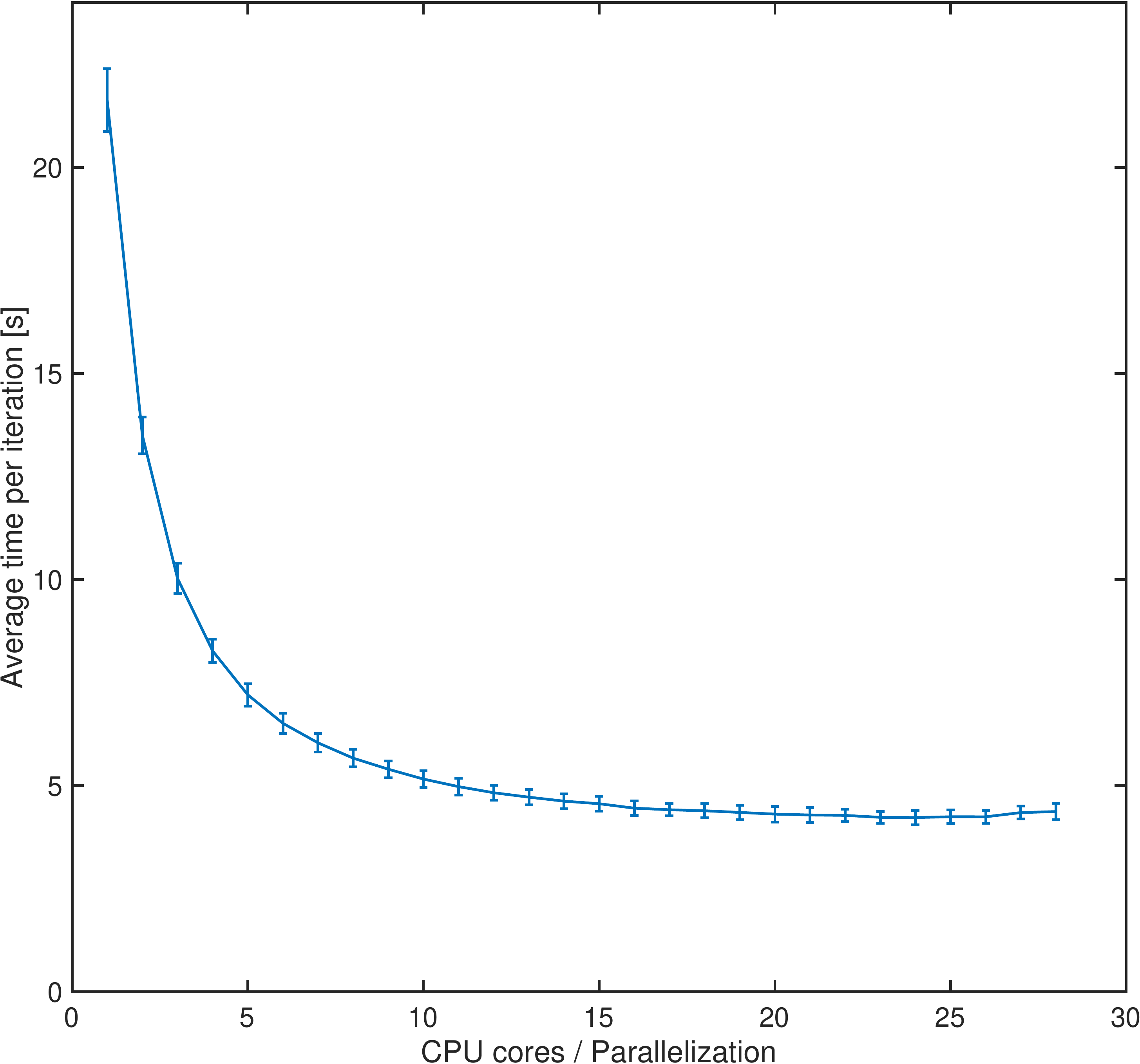}
 \caption{The EXACT gradient computation times (mean $\pm$ standard deviation (errorbars))  were estimated by optimizing five different cases limited to ten iterations. Each optimization was tested with up to 28-fold parallelization. }%
 \label{FigSTPI}
 \end{figure*}

\pagebreak

\section{Flip Angle Maps}

\begin{figure*}[ht!]
\includegraphics[width=\textwidth, angle=0]{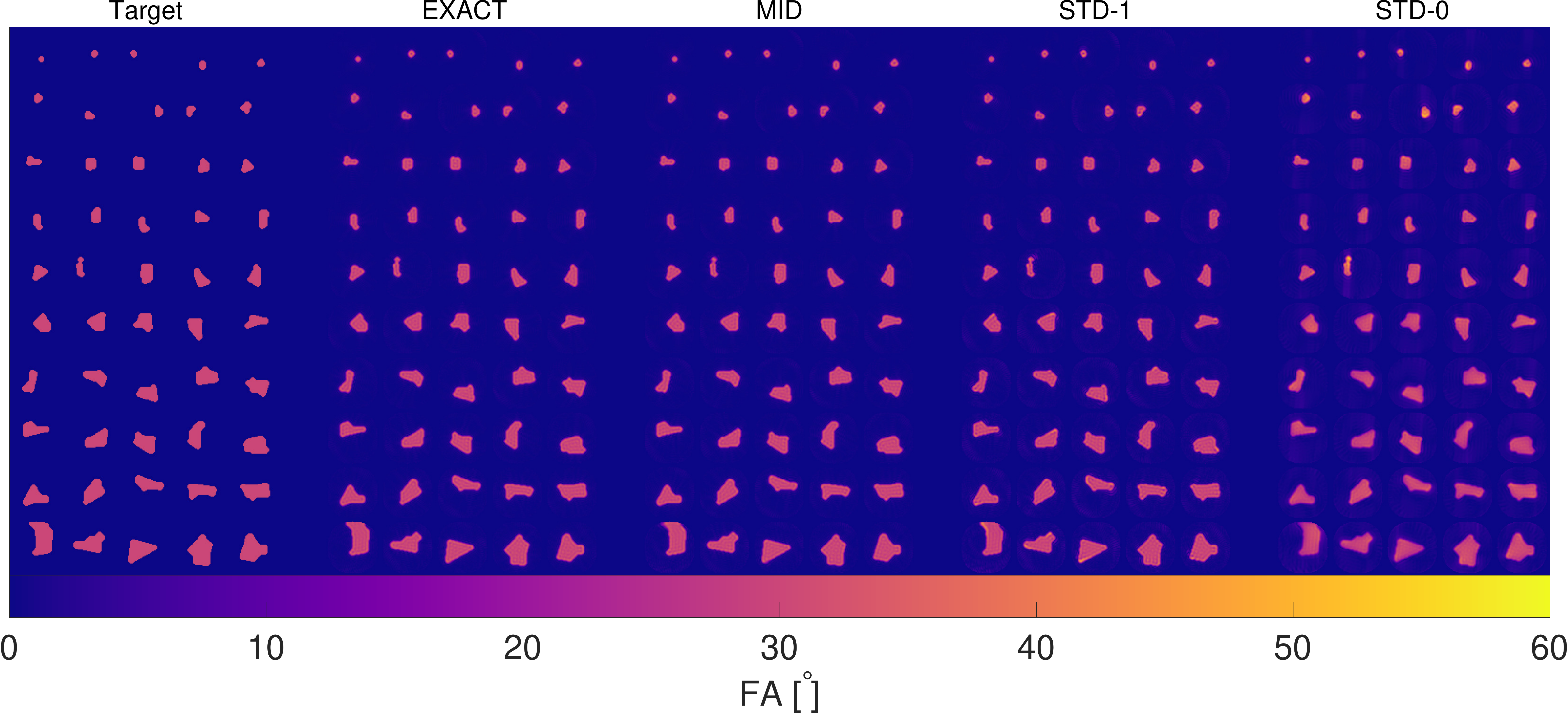}
\caption{Flip-angle (FA) maps in the BW (black-white) category. From left to right: Target maps (ten points were randomly placed inside the mask depicted in Figure 3, dilated and filled to form single arbitrary shape); FA maps produced by the EXACT, MID, STD-1 and STD-0 gradient methods, respectively.
}%
\label{FigSFABW}
\end{figure*}

\begin{figure*}[ht!]
\includegraphics[width=\textwidth, angle=0]{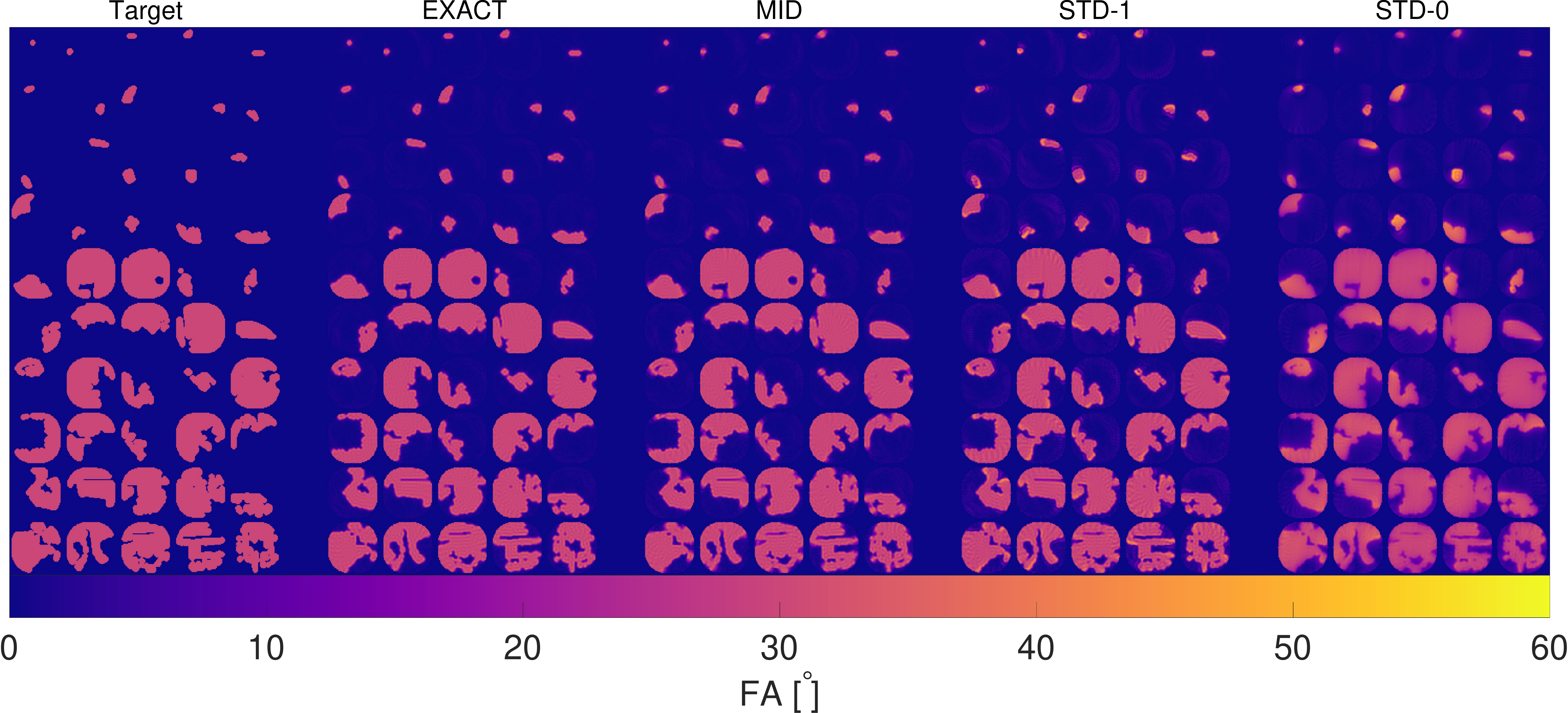}
\caption{Flip-angle (FA) maps in the GrBW category.  From left to right: Target maps (Gray-scale images were normalized and 0.5-thresholded, and scaled flip-angle wise by the nominal FA of $30^\circ$. The super-ellipsis mask was then applied); FA maps produced by the EXACT, MID, STD-1 and STD-0 gradient methods, respectively.
}%
\label{FigSFAGrBW}
\end{figure*}

\begin{figure*}[ht!]
\includegraphics[width=\textwidth, angle=0]{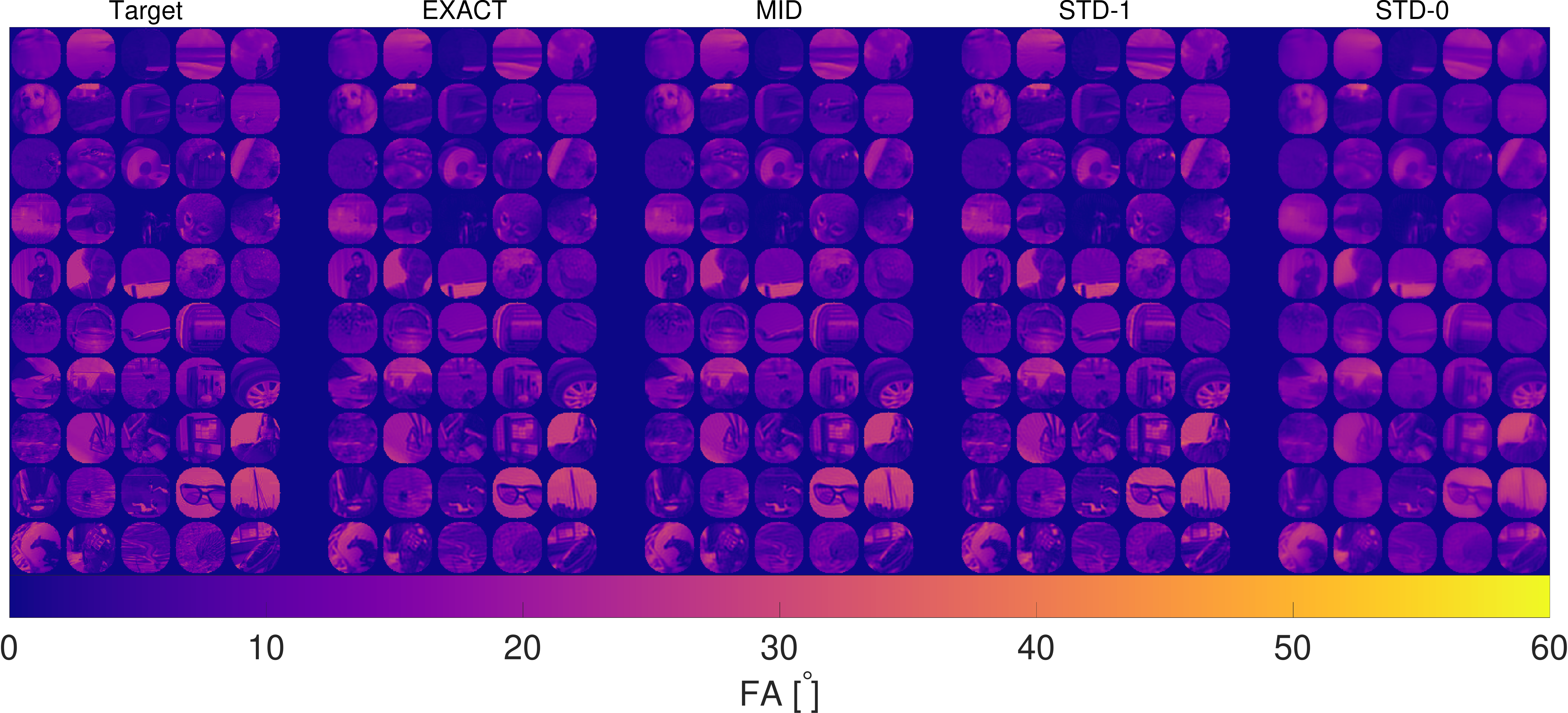}
\caption{Flip-angle (FA) maps in the Gr category. From left to right: Target FA maps (Gray-scale images were normalized and scaled flip-angle wise by the nominal FA of $30^\circ$. The super-ellipsis mask was then applied); FA maps produced by the EXACT, MID, STD-1 and STD-0 gradient methods, respectively.
}%
\label{FigSFAGr}
\end{figure*}

The flip angle (FA) maps displayed in this sections, Figures \ref{FigSFABW}, \ref{FigSFAGrBW}, and \ref{FigSFAGr}, are all sorted after increasing NRMSE of the EXACT method; row-wise, with lowest NRMSE in the top left corner. 

The two standard gradient methods tend to produce FA's beyond the nominal FA of $30^\circ$ inside the target regions, and have a worse tendency to elevate FA's beyond $0^\circ$ outside the target region.

\end{document}